\documentclass[letterpaper, 10pt, conference]{ieeeconf}
\IEEEoverridecommandlockouts  
\usepackage{amsmath}

\usepackage{graphicx}

\usepackage[colorlinks=true, allcolors=blue]{hyperref}

\usepackage{tabularx}
\usepackage{booktabs}
\usepackage{makecell}
\usepackage{array}
\newcolumntype{Y}{>{\raggedright\arraybackslash}X}

\title{\LARGE \bf
The Impact of Privacy and Security Attitudes and Concerns of Travellers on Their Willingness to Use Mobility-as-a-Service Systems}
\author{Maria Sophia Heering, Haiyue Yuan, Shujun Li\\
Institute of Cyber Security for Society (iCSS) \& School of Computing, University of Kent, UK\\
\{m.s.heering, h.yuan-221, s.j.li\}@kent.ac.uk%
\thanks{This is the authors' version of the accepted paper. Please cite this paper as follows: Maria Sophia Heering, Haiyue Yuan and Shujun Li (2023) The Impact of Privacy and Security Attitudes and Concerns of Travellers on Their Willingness to Use Mobility-as-a-Service Systems. \emph{Proceedings of the 2023 IEEE 26th International Conference on Intelligent Transportation Systems} (ITSC 2023), pp.~5573--5578, IEEE, doi: \href{https://doi.org/10.1109/ITSC57777.2023.10422468}{10.1109/ITSC57777.2023.10422468}. For the published version, please visit the publisher's website via the DOI link.}%
\thanks{This work was supported by the Engineering and Physical Sciences Research Council (EPSRC), part of the UK Research and Innovation (UKRI), under grant numbers EP/V039164/1}}

\begin{document}
\maketitle

\begin{abstract}
This paper reports results from an online survey on the impact of travellers' privacy and security attitudes and concerns on their willingness to use mobility-as-a-service (MaaS) systems. This study is part of a larger project that aims at investigating barriers to potential MaaS uptake. The online survey was designed to cover data privacy and security attitudes and concerns as well as a variety of socio-psychological and socio-demographic variables associated with travellers' intentions to use MaaS systems. The study involved $n=320$ UK participants recruited via the Prolific survey platform. Overall, correlation analysis and a multiple regression model indicated that, neither attitudes nor concerns of participants over the privacy and security of personal data would significantly impact their decisions to use MaaS systems, which was an unexpected result, however, their trust in (commercial and governmental) websites would. Another surprising result is that, having been a victim of improper invasion of privacy did not appear to affect individuals' intentions to use MaaS systems, whereas frequency with which one heard about misuse of personal data did. Implications of the results and future directions are also discussed, e.g., MaaS providers are encouraged to work on improving the trustworthiness of their corporate image.
\end{abstract}

\section{Introduction}
\label{sec:introduction}

With a worldwide growing population, increases in urbanisation levels and associated growing concerns for environmental issues, the transport sector finds itself in a crucial position and in need of more modern, environmentally sustainable and efficient solutions. In fact, transport is recognised as one of the sectors with the largest greenhouse gas emissions in many countries and worldwide~\cite{Sulskyte2021}. Providing more efficient and greener mobility solutions in urban, suburban and rural areas could produce personal benefits for individuals (e.g., long-term reduced costs as a result of not owning any personal vehicles, more travel choices, and a healthier life style) as well as wider benefits for the society and the planet as a whole (e.g., reduced traffic congestion, and greenhouse gas emissions with a consequent lower impact on global warming). The number of empirical and theoretical studies supporting the development of green transport has been rapidly growing with the aim to promote more sustainable societies, improve individuals' quality of life and creating more functional travel solutions.

Mobility-as-a-service (MaaS), a multi-modal transport service which offers passengers with seamless and end-to-end mobility options, appears to be a significant step in the direction of more efficient and environmentally friendly transport. MaaS offers an integrated system that allows travellers to plan, book and pay for traditional services, such as public transport, as well as on-demand and shared services (e.g., ride-, bike- and car-share) via a single platform~\cite{Casado2020, Liimatainen2021}. The development and large-scale uptake of this service is expected to reduce private vehicle usage with consequent positive effects on traffic congestion and air and acoustic pollution~\cite{Eckhardt2020}. To work efficiently, MaaS requires a complex infrastructure and an efficient digital network of stakeholders~\cite{Chinaei2022}. However, even when these are provided, a successful implementation is not automatically guaranteed. In fact, the potential of MaaS largely depends on the willingness of travellers to accept these technologies and to change their travelling behaviours and habits accordingly. In order to successfully implement MaaS, it is thus necessary to explore travellers' attitudes, worries and needs~\cite{Schikofsky2020, Alonso-GonzalezZ2020}. Among the worries that potential users might have, and which could work as barriers to the adoption of MaaS, are users' concerns over the privacy and security of their personal data~\cite{Butler2021, Ye2020}. In fact, the complex and integrated system required for MaaS to coordinate multi-modal solutions relies on the integration of diverse transport service providers and related stakeholders (e.g., payment processors), who acquire, exchange and process operational data, mostly in a decentralised manner. The decentralisation and flexibility comes with many security and privacy risks~\cite{Chinaei2022, Callegati2018, Bothos2019}.

Previous research has considered the privacy and security of personal data as potential barriers to the implementation of MaaS, which however have been discussed mostly from a system-security perspective~\cite{Chinaei2022, Callegati2018, Bothos2019, Kong2021, Hong2013} or a policy and regulations perspective~\cite{Liimatainen2021, Kong2021, Jittrapirom2018, Audouin2018}. As we will discuss in Section~\ref{sec:privacy_security_concerns}, how attitudes and concerns of travellers about the privacy and security of their personal data could affect their willingness to use MaaS systems is still less studied with scarce and inconclusive evidence~\cite{Schikofsky2020, Alonso-GonzalezZ2020, Ye2020, Polydoropoulou2020, Caiati2020, Becker2021}. We aim to contribute to this literature by looking at several indicators of privacy and security perceptions and how these relate to travellers' intention to use MaaS systems. To this end, our survey combines several existing scales on internet privacy and security attitudes and concerns (see Section~\ref{sec:method}). Among the indicators of privacy and security concerns, we also included trust in both commercial and governmental websites~\cite{Hong2013}, which are potentially `more relatable' indicators. Additionally, we also looked at indicators of personal experiences of internet misuse and familiarity with news about misuse of personal data. Analysing data from our survey (see Section~\ref{sec:results}), we were able to answer the following research questions:
\begin{itemize}
\item \textbf{RQ1}: Do cyber security and privacy concerns and attitudes affect travellers' decisions to use MaaS systems?

\item \textbf{RQ2}: Does having been a victim of perceived improper invasion of privacy have an impact on travellers' decisions to use MaaS systems?

\item \textbf{RQ3}: Does the frequency with which a traveller has come across news of potential misuse of personal data affect their decisions to use MaaS systems?

\item \textbf{RQ4}: Does trust in how websites handle users' personal data affect their decisions to use MaaS systems?
\end{itemize}

This research contributes to the literature by concluding that the role that travellers' internet privacy and security concerns have on their willingness to use MaaS systems is more related to `trust in the provider' and the frequency with which people have come across news about misuse of personal data. This gives rise to several recommendations for MaaS researchers and providers (see Section~\ref{sec:conclusions}).

\section{Related Work}
\label{sec:related_work}

Privacy and security risks have been recognised as critical aspects of the development of MaaS. Research has looked at both technical and socio-technical security risks. We consider a risk technical if it is more about technologies used in MaaS, e.g., denial-of-service (DoS) attacks, ransomware attack~\cite{Callegati2018, Mouhibbi2018, Nguyen2019}, and a risk socio-technical if it is more related to socio-technical factors such as behaviours of travellers, operators of MaaS and other related stakeholders, and policy makers (e.g., data misuse through profiling and inference, `unruly' third-party access and industrial espionage). Although the importance that privacy and security risks have in the development of MaaS has been acknowledged, there is still insufficient research considering the role that travellers' privacy and security attitudes and concerns have on their willingness to use MaaS systems~\cite{Schikofsky2020, Alonso-GonzalezZ2020, Ye2020, Polydoropoulou2020, Caiati2020, Becker2021}. In fact, while some studies concluded that privacy and security concerns could negatively affect travellers' decisions to adopt MaaS~\cite{Ye2020, Alonso-GonzalezZ2020}, some found that these concerns had no impact~\cite{Schikofsky2020, Caiati2020} and others reported mixed results~\cite{Polydoropoulou2020, Caiati2020}. Furthermore, past studies have seldom considered the role played by trust in the providers~\cite{Casado2020, Jung2020, Vij2020} and, to the best of our knowledge, no study has simultaneously measured the impact of privacy and security attitudes/concerns and trust in the providers on travellers' willingness to use MaaS systems. Our work therefore fills this gap, in addition to also looking at two other factors (having been a victim of information misuse and the frequency with which one has heard of information misuse) that may affect travellers' decisions to use MaaS systems.

\subsection{Privacy and Security Concerns}
\label{sec:privacy_security_concerns}

Privacy and security concerns and risks associated with MaaS systems are closely related to the types of personal data collected and how the collected personal data are used and shared. A number of studies have looked into how personal data collected by MaaS systems can be explored to infer user's behaviour and mobility patterns~\cite{Barreto2018, Costantini2019}. For instance, MaaS users' movement data can be analysed to infer information about certain health conditions~\cite{Costantini2019}; location data with information about time of use can be monetised by companies, introducing potential privacy and ethical concerns~\cite{Cooper2019}; a driver's performance data and GPS coordinates can reveal sensitive information, leading to violations of their identity and location privacy~\cite{Belletti2017, Kong2021}. Privacy concerns in MaaS systems extend beyond profiling and inference risks to include third-party access to personal data and over-sharing of personal data between multiple parties. Different stakeholders require access to and processing of personal data for a MaaS system to function effectively. It is essential to examine third-party processors such as payment processors and hosting providers to address privacy considerations and implications~\cite{Cottrill2020}. It is also important to understand what data are necessary for what operations and what data are requested unnecessarily by stakeholders~\cite{Pitera2017}. Similarly, some research recommended that users should be given a certain degree of privacy control when their personal data are shared on open data platforms, with the consensus that personal data types and formats can be shared to promote the smooth operation of mobility platforms~\cite{He2021}. To overcome these obstacles, it was highlighted in a previous study that privacy regulations should be carefully considered for supporting the development of MaaS and for enhancing trust of both users and providers~\cite{Butler2021}.

\subsection{Trust}
\label{sec:trust}

Trust has been argued to be an important prerequisite for successful e-commerce and e-services. This is because online transactions occur with a high degree of uncertainty (admittedly higher than in face-to-face exchanges) with transactions being blind, borderless and non-instantaneous, online users need to trust the sellers and providers that they will fulfil their obligations without engaging in harmful behaviours (e.g., providing inaccurate information, violating the customers' privacy, and making unauthorised use of credit card information)~\cite{Kim2008, Gefen2003}. A multitude of studies have indeed shown that trust plays an essential role in online transactions, both directly and indirectly through the reduction of consumers' perceived risk~\cite{Kim2008, Gefen2003, Qalati2021}.

Although trust of travellers has been widely considered as having a crucial role in the implementation of MaaS systems~\cite{Butler2021, Jung2020}, its impact on travellers' willingness to use MaaS systems has been seldom studied. This is surprising considering the following aspects: a) there is a vast literature on the role that trust plays on consumers' acceptance of e-sellers and e-services, b) research on MaaS has shown that trust in the MaaS providers is positively associated with intentions to purchase MaaS bundles~\cite{Vij2020}, and c) research on MaaS has shown that individuals would wish for the government (admittedly a respectful and trustful stakeholder) to play an active role in the service (both as an overseer or a provider)~\cite{Casado2020,Jung2020, Vij2020}. We consider the low number of studies assessing and measuring the impact that trust in the MaaS providers and its impact on travellers' intention to use MaaS as an important gap in the literature. More specifically, we feel the need to concurrently measure users' trust in how the provider handles their personal data and their privacy and security concerns. In fact, we believe that trust could be a useful proxy for privacy and security concerns, as it being potentially more familiar and relatable for most non-expert users with less knowledge on privacy and security matters. 

\section{Methodology}
\label{sec:method}

\subsection{Procedures and Materials} 

There are a number of past studies that investigated the impact of privacy and security attitudes and concerns on people's willingness to use MaaS systems. However, such studies are often fragmented and did not systematically consider different aspects of MaaS systems from end users' perspectives. We conducted a large-scale online survey, aiming to comprehensively learn about travellers' security and privacy attitudes in relation to their willingness to use MaaS systems. We included several existing scales on internet privacy and security attitudes and concerns in order to test their potentially different impact on travellers willingness to adopt MaaS. In line with the gaps identified (see Section~\ref{sec:trust}) we also included measures of trust in commercial and governmental websites. Additionally, because trust in and reputation of a provider can be affected by social-environment cues like media stories of hacking and loss of credit card details~\cite{Morton2014}, we decided to measure how much participants had heard/read about the use of potential misuse of information collected on the Internet and assess whether this would impact the decision to use MaaS systems. Based on previous literature we expected familiarity with information misuse to negatively affect the willingness to use MaaS~\cite{Morton2014} and for personal experiences of information misuse to do the same. The survey also included a variety of socio-psychological and travel-behaviour related variables that will be discussed in a follow-up paper. The online survey was designed to consist of the following six parts.

\begin{itemize}
\item Part 1 contains socio-demographic questions.

\item Part 2 focuses on \textit{transport and travel route information behaviour}. Participants were asked questions about their transport and travel information habits (e.g., what trip and route information apps they use and how often they use such apps). 

\item Part 3 focuses on \textit{data sharing}. Participants were asked questions on how they feel about sharing personal data online (i.e., internet privacy concerns) and on how much they trust or are concerned with the way commercial and governmental websites dispose of users' personal data.

\item Part 4 primarily investigates \textit{perceptions about MaaS systems}. Participants were presented with a brief description of MaaS and were subsequently asked how they felt about the service (e.g., perceived usefulness of MaaS, intention to use MaaS if it were available, perceived incentives to use MaaS).

\item Part 5 looks at \textit{transportation habits and evaluations}. Participants were asked questions aimed at exploring their personal transport habits and their experiences and impressions of their local public transport systems (e.g., what is their primary mode of transportation, their level of satisfaction with local buses).

\item Part 6 is included to learn about how often if ever participants had became a victim of improper invasion of privacy online and how often if ever they had heard about news on misuse of personal data.
\end{itemize}
After the above six parts, participants were invited to provide any further comments, thanked and debriefed.

This study received a favourable ethical opinion from the Central Research Ethics Advisory Group of the University of Kent (Reference Number: CREAG109-09-22). We used the Jisc Online Surveys system\footnote{\url{https://www.onlinesurveys.ac.uk/}} to host the online survey and the crowdsourcing platform Prolific\footnote{\url{https://www.prolific.co/}} to recruit participants. All participants gave their consent electronically as part of the online survey before proceeding to take the online survey. The participants were compensated financially at a rate of £9 per hour and the survey took an average participant 17 minutes to complete. We carried out a power analysis using G*Power~\cite{Faul2007}, suggesting that a sample size of at least 274 participants is needed to detect a small-to-medium effect size, $f = .17$, at 80\% power ($\alpha = .05$). To be on the side of caution, we decided to recruit 320 participants.

\subsection{Measures} 
\label{sec:measures}

It is worth noting that the emphasis of this paper is on data privacy and security concerns of MaaS, we hereby focus on investigating 1) variables from Part 3 of the survey on data sharing and their relationships with the behavioural intention to use MaaS systems; and 2) variables from Part 6 of our survey, relative to experiences of invasion of privacy and news of information misuse. Variables from the remaining parts of the survey (parts 1, 4 and 5) will be analysed in a follow-up paper. We computed bi-variate correlations to identify which of our variables were significantly associated with the`behavioural intention to use MaaS'. We then used a multiple regression analysis, with `behavioural intention to use MaaS' as our primary dependent variable (DV), to identify which of those variables would work as a significant predictor (IVs) of the DV. More details about all variables used in this study are presented as follows.

\subsubsection{DV: Behavioural Intention to Use MaaS (BIUM)}

The intention to use MaaS was measured by asking participants to indicate how much they disagreed or agreed with the following statements using a 7-point Likert scale (1 = strongly disagree, 7 = strongly agree; $\alpha = .97$):
\begin{itemize}
\item `\emph{Assuming I would have access to the MaaS offering, I intend to use it.}'

\item `\emph{I expect to use the MaaS offering when it becomes available.}'

\item `\emph{Given that I would have access to the MaaS offering, I predict using it.}'
\end{itemize}


\subsubsection{IV: Attitudes towards Personal Identifying Information collection (APII)}

Participants were asked to indicate how much they disagreed or agreed with seven different statements borrowed from a previous study~\cite{Becker2021}. Examples of statements are: `\emph{I want a website to disclose how my PII will be used}', `\emph{I am unconcerned when a website uses my PII to customise my browsing experience (R)}' and `\emph{I mind when a website that I visit collects (without my consent) information about my browser configuration}' (1 = strongly disagree, 7 = strongly agree; $\alpha = .84$).

\subsubsection{IVs: Internet Privacy Concerns for Commercial Websites (IPCC) \& Internet Privacy Concerns for Governmental Websites (IPCG)} 

To assess participants' internet privacy concerns we asked them about their level of agreement with 18 different statements borrowed from a previous study~\cite{Kong2021}. Reliability of this scale was very high, respectively $\alpha = .96$ for commercial websites and $\alpha = .97$ for governmental websites. These items refer to six different domains (Collection, Secondary Usage, Errors, Improper Access, Control and Awareness) and are considered separately for \emph{commercial} and \emph{governmental} websites.

\subsubsection{IVs: Trusting Beliefs for Commercial Websites (TBC) \& Trusting Beliefs for Governmental Websites (TBG)}

We asked participants to indicate their level of agreement with four different statements borrowed from a previous study~\cite{Hong2013}, with the purpose of assessing how much individuals believe they can trust \emph{commercial} and \emph{governmental} websites on handling their personal data. Two example statements are: 
\begin{itemize}
\item `\emph{Commercial/Governmental websites in general would be trustworthy in handling my personal information.}'

\item `\emph{Commercial/Governmental websites would fulfil their promises related to my personal information.}'
\end{itemize}
All these questions are measured using a 7-point Likert scale (1 = strongly disagree, 7 = strongly agree; $\alpha 
 = .93$ for trust related to commercial websites and $\alpha = .95$ for trust related to governmental websites).


\subsubsection{IV: Improper Invasion of Privacy (IIP)}

Participants were asked one question: `\emph{How frequently have you personally been the victim of what you felt was an improper invasion of privacy?}' (1 = Never, 6 = Frequently).

\subsubsection{IV: News of Information Misuse (NIM)}

Participants were asked: `\emph{How much have you heard or read during the last year about the use and potential misuse of the information collected from the Internet?}' (1 = Not at all, 7 = Very much).

\section{Results}
\label{sec:results}

320 individuals from the UK (133 females, 183 males, 3 preferred not say, 1 other) took part in this study. The mean age was 39.71 ($\text{SD} = 11.79$). The majority of the participants reported to be White (85.9\%), followed by Asian or Asian British (10\%), Black/Black British/ African or Caribbean (2.2\%), and (0.9\%) from mixed or other ethnicity. No participants were excluded or included following data analysis.

\begin{table*}[!htb]
\centering
\caption{Means, Standard Deviations (SD) and Correlations among Variables}
\label{tab:stats}
\begin{tabular}{*{13}{c}}
\toprule
Measures & Mean & SD & BIUM & APII & IPCC & IPCG & TBC & TBG & IIP & NIM\\
\midrule
BIUM & 4.67 & 1.52 & - & \makecell{-.06\\($p = .27$)} & \makecell{-.03\\($p = .57$)} & \makecell{-.07\\($p = .21$)} & \makecell{.24\\($p \le .001$)} & \makecell{.18\\($p \le .001$)} & \makecell{.04\\($p = .50$)} & \makecell{.15\\($p = .008$)}\\
\midrule
APII & 5.44 & 1.05 & & - & \makecell{.77\\($p = .008$)} & \makecell{.48\\($p = .008$)} & \makecell{-.34\\($p = .008$)} & \makecell{-.24\\($p = .008$)} & \makecell{.33\\($p = .008$)} & \makecell{.27\\($p = .008$)}\\
\midrule
IPCC & 5.48 & 1.02 & & & - & \makecell{.56\\($p \le .001$)} & \makecell{-.36\\($p \le .001$)} & \makecell{-.22\\($p \le .001$)} & \makecell{.35\\($p \le .001$)} & \makecell{.30\\($p \le .001$)}\\
\midrule
IPCG & 4.25 & 1.37 & & & & - & \makecell{-.29\\($p \le .001$)} & \makecell{-.55\\($p \le .001$)} & \makecell{.33\\($p \le .001$)} & \makecell{.31\\($p \le .001$)}\\
\midrule
TBC & 3.80 & 1.37 & & & & & - & \makecell{.52\\($p \le .001$)} & \makecell{-.25\\($p \le .001$)} & \makecell{-.12\\($p = .03$)}\\
\midrule
TBG & 4.77 & 1.38 & & & & & & - & \makecell{-.17\\($p \le .001$)} & \makecell{-.14\\($p = .01$)}\\
\midrule
IIP & 2.31 & 1.08 & & & & & & & - & \makecell{.25\\($p \le .001$)}\\
\midrule
NIM & 4.26 & 1.51 & & & & & & & & -\\
\bottomrule
\end{tabular}
\end{table*}

\subsection{Analysis of Correlations}
\label{sec:correlation}

First of all, we looked at the correlations between `behavioural intentions to use MaaS' (i.e., DV: BIUM) and indexes for data privacy and data security attitudes and concerns. As suggested by the authors who introduced the scales for Internet Privacy Concerns~\cite{Hong2013}, we computed separate indexes for commercial and government websites (i.e., IVs: IPCC \& IPCG). As shown in Table~\ref{tab:stats}, participants had moderate to high levels of privacy and security concerns on their personal data (IPCC: $M = 5.48$, $\text{SD} = 1.02$; IPCG: $M = 4.25$, $\text{SD} = 1.37$; and APII: $M = 5.44$, $\text{SD} = 1.05$). However, interestingly and somewhat surprisingly (somewhat because evidence in literature is mixed and so far inconclusive), we found that none of these indexes were significantly correlated with participants' intentions to use MaaS (see Table~\ref{tab:stats}). This mismatch between the level of concern and the absence of a relationship between the concern and the intention to use MaaS appears to be in line with literature on the `privacy paradox', which suggests that, although individuals tend to indicate privacy as a primary concern, they reveal personal information for relatively small rewards~\cite{Kokolakis2017}.

Contrastingly, correlations between BIUM and TBC ($r = .24$, $p \le .001$) and between BIUM and TBG ($r = .18$, $p \le .001$) are both positive and significant (although weak), indicating that the trust on commercial and government websites to handle personal data plays a role on nudging participants' intention to use MaaS. Additionally, BIUM was also positively correlated with NIM ($r = .15$, $p = .01$), suggesting that participants' intention to use MaaS is associated with the frequency with which participants had heard about information misuse on the internet. Whereas, there is no correlation between BIUM and IIP ($r = .04$, $p = .50$), which reveals that participants' past experiences of personal invasion of privacy does not seem to affect their willingness to use MaaS. This last result may be surprising, since one could reasonably expect that past experiences of personal invasion of privacy would have a negative impact on people's willingness to use other apps in the future. One plausible explanation for the lack of this correlation is related to the distribution of data. People generally believe that they only infrequently have been victims of misuse of information collected from the internet ($M = 2.31$, $\text{SD} = 1.08$), suggesting that it could be difficult to detect a significant relationship. Finally, and coherent with the notion that to higher trust should correspond lower concerns of users, the correlations between the three indicators of privacy and security attitudes/concerns (i.e., APII, IPCC, and IPCG) and trust in commercial/government websites (i.e., TBC and TBG) are all negative and significant (see Table~\ref{tab:stats}).

\subsection{Regression Analysis}
\label{sec:regression}

By considering the results from the correlation analysis presented in Section~\ref{sec:correlation}, here we present a multiple linear regression model to investigate the effects of those variables (i.e., TBC, TBG, and NIM) that have significant correlations with BIUM. 
The fitted regression model was ($p \le.001$): 
\begin{equation*}
\text{BIUM} = 2.46 + .23\times\text{TBC} + .11\times\text{TBG} + .19\times\text{NIM}.
\end{equation*}

The overall regression was statistically significant ($R = 0.31$, $F(3, 316) = 11.09$, $p \le .001$, $R^2 = .095$, $R^2\text{ Adjusted} = .087$), however, the model only explains a small amount of the variance in the value of intentions to use MaaS (9.5\%). Additionally, it was found that, whereas `trust in commercial websites' (i.e., TBC) ($\beta = 0.20$, $p \le.001$) and `frequency of news misuse' (i.e., NIM) ($\beta = 0.11$, $p \le .001$) significantly predicts `intentions to use MaaS' (i.e., BIUM), trust in governmental websites' (i.e., TBG) does not ($\beta = 0.10$, $p = 0.11$). 

\subsection{Results and Discussion}
\label{sec:results_discussion}

Although MaaS has received increasing interest since it was first presented at ITS Europe Congress held in Helsinki, Finland, in June 2014~\cite{Sakai2019}, research investigating the effects of privacy and security concerns on the intentions to use MaaS by travellers, is still scarce and so far inconclusive. We aimed to bring some clarity to this area of research by concurrently measuring users’ trust in how
the provider handles their personal data and their privacy and security attitudes and concerns. In our study we did not find any relationship between data privacy/security attitudes/concerns and MaaS usage, however, we found that trust, and more specifically trust in a (commercial and governmental) website's handling of users' personal data, does positively predict our participants' intentions to use MaaS. This result is in line with previous research on MaaS, which identified trust in the provider as relevant and a positive predictor of willingness to adopt MaaS~\cite{Casado2020, Jung2020, Vij2020}, and more importantly, this result is in line with a long tradition of research that identifies trust on website as a core positive predictor of people's intentions to use or buy from online service providers.

One limitation of this study is that we measured trust as trust in how general commercial/governmental websites handle users' personal information and not as trust in how a MaaS system would handle users' personal information. However, we argue that because we did not use a stated preference approach (where participants are asked to make decisions in hypothetical choice scenarios) but simply presented participants with a generic definition of a MaaS system, it would have been too artificial to ask participants how much they would trust this hypothetical system. Additionally, because MaaS systems will necessarily be considered as either commercial or governmental (or potentially hybrid) systems, the trust measures we used should be relevant to MaaS as well.

\section{Conclusions}
\label{sec:conclusions}

In this survey-based study, we investigated the role that privacy and security concerns have on travellers' willingness to use MaaS systems. We concurrently (to the best of our know, for the first time in the literature on MaaS) assessed the role played by trust in the providers and the role played by users' privacy and security attitudes and concerns. Additionally, we tested whether having been a victim of improper invasion of privacy and the frequency with which individuals had heard/read of episodes of internet information misuse would affect their willingness to use MaaS. We did not find evidence to show that privacy and security concerns has significant impact on our participants' intention to use MaaS. Contrastingly, we found that trust in how users' data is handled, and more specifically trust in how commercial and governmental websites handle users' data, does have an (although small) positive impact on participants' intention to use MaaS. These results are in line with the vast literature showing how trust plays a pivotal role in consumers' acceptance of e-sellers and e-services. We would recommend that \emph{research investigating the role that privacy and security concerns have on travellers' willingness to use MaaS should also consider their trust on providers' handling of personal data}. In comparison with privacy and security concerns, trust could in fact be a more relatable concept for potential MaaS users and a clearer predictor.

This work furthers our understanding of the role played by privacy and security concerns on travellers' intention to use MaaS systems and directs the attention towards the significant influence that trust on providers can have on the MaaS usage intention. We suggest that, \emph{as for most online services, in order to increase the number of customers and to decrease users' perceived risks, MaaS providers must work to build a more trusted and reliable image among people}. 
For example, to decrease users' perceived privacy and security risks, MaaS systems could provide assurance that they comply with a privacy policy that clearly indicates what personal data will be collected and how such collected personal data will be used and shared~\cite{Kim2008}.

\bibliographystyle{IEEEtran}
\bibliography{main.bib}

\end{document}